\documentclass[journal]{IEEEtran}
\usepackage{amsmath,amssymb,amsfonts}
\usepackage{algorithmic}
\usepackage{graphicx}
\usepackage{textcomp}
\usepackage{xcolor}
\usepackage{float}
\usepackage{diagbox}
\usepackage{booktabs}
\usepackage{multirow}
\usepackage{multirow}
\usepackage[table]{xcolor}

\usepackage[
backend=biber,
style=ieee,
sorting=none
]{biblatex}
\def\BibTeX{{\rm B\kern-.05em{\sc i\kern-.025em b}\kern-.08em
    T\kern-.1667em\lower.7ex\hbox{E}\kern-.125emX}}
\begin{document}

\title{Dynamic and Open-Set RF Fingerprinting and Localization in Crowded Indoor Environments through Contrastive Channel State Information Learning \\
\thanks{This work was funded by U.S. National Science Foundation (NSF) under grant 2233774.}
}

\author{Fawaz Abdul Razak, Yasin Yilmaz%

\thanks{Fawaz Abdul Razak is with Department of Electrical Engineering , University of South Florida, Tampa, United States (e-mail: fawaz243@usf.edu)}

\thanks{Yasin Yilmaz  is with Department of Electrical Engineering , University of South Florida, Tampa, United States (e-mail: yasiny@usf.edu)}}

\renewcommand{\vec}[1]{\boldsymbol{#1}}
\newcommand{\vr}{\vec{r}}
\newcommand{\vx}{\vec{x}}
\newcommand{\vy}{\vec{y}}
\newcommand{\vz}{\vec{z}}

\maketitle


\begin{abstract}
Radio Frequency Fingerprinting (RFF) using deep learning has gained attention as a complementary approach to cryptographic authentication, offering resistance to spoofing, replay attacks, and key leakage. While most RFF approaches rely on In-Phase and Quadrature (IQ) samples, Channel State Information (CSI) has emerged as a more accessible alternative, enabling device authentication through physical-layer characteristics. In this work, we propose \textit{ContraCSI}, a CSI-based contrastive learning framework for RFF using low-cost ESP32 devices. We investigate multiple encoder backbones, including a Vision Transformer (ViT), a lightweight 3D-CNN (Lite3D-CNN), and R3D18, to learn joint CSI and device-ID embeddings for transmitter authentication. For closed-set identification, the ViT variants achieve the best overall performance. We further study \emph{open-set} authentication by applying a Geometric Entropy Minimization (GEM)-based anomaly score and sequential CUSUM (Cumulative Sum) test on embeddings learned by Lite3D-CNN-Contra, enabling rejection of unseen or non-enrolled transmitters rather than forcing a closed-set label. To evaluate robustness in highly dynamic and crowded indoor environments with human motion, multipath fading, and varying device orientations and distances, we conduct extensive experiments in a real-world setting. Our results demonstrate high authentication accuracy, strong generalization in non-ideal conditions, and effective rejection of unknown transmitters. Additionally, we explore CSI-based indoor localization via trilateration, illustrating the potential for integrated authentication and localization in practical indoor deployments.
\end{abstract}


\section{Introduction}

Radio Frequency Fingerprinting (RFF) has gained significant prominence due to the surge in IoT devices being manufactured and deployed every year. RFF-based authentication aims to mitigate the vulnerabilities of cryptographic methods, which are susceptible to key extraction, cloning, and spoofing attacks, and require computational and storage resources often limited in lightweight IoT devices. Unlike key-based authentication, RFF relies on unavoidable hardware imperfections in transmitters to serve as device identifiers, which are physically impossible to replicate.

With the success of deep learning (DL) models such as Convolutional Neural Networks (CNNs) and Transformers in multidimensional feature classification, RFF-based authentication has seen significantly improved performance. Most DL models for RFF process raw IQ samples, which provide RF fingerprints with minimal noise distortion. However, collecting raw IQ samples requires specialized equipment such as Software Defined Radios (SDRs) or Network Interface Cards (NICs). Channel State Information (CSI), which carries device-specific distortions (see Sec.~\ref{sec:proposed_sys}), on the other hand, is computed during every WiFi packet transmission, and can be easily accessed using open-source toolkits on low-cost microcontrollers such as the ESP32.

\textbf{\textit{Static vs.\ Dynamic RFF:}}
A critical but often overlooked distinction in the RFF literature is between \emph{static} and \emph{dynamic} evaluation settings. In a static setting, transmitters and receivers remain at fixed positions during both training and testing. Under this condition, the estimated CSI, $\hat{H}_k = h_k \cdot \psi_k + \varepsilon_k$, is dominated by a nearly constant channel factor $h_k$, which encodes the specific multipath profile of that fixed geometry. Consequently, the learned fingerprint is a composite of the true device signature $\psi_k$ and the position-dependent channel response $h_k$. These \emph{position-dependent signatures} are not true device fingerprints: they change as soon as the transmitter or receiver is relocated, even by a small distance, because the multipath structure and hence $h_k$ changes. Models trained in static conditions therefore tend to learn spatial signatures rather than genuine hardware impairments, leading to a false sense of high accuracy that does not generalize. Only in a \emph{dynamic} setting, where transmitter--receiver geometry varies across data collection sessions, the model is forced to learn features that are invariant to channel variations and truly reflect hardware-level impairments $\psi_k$. This is why dynamic-channel evaluation is essential for practical RFF deployment.

\textbf{\textit{Research Gap:}}
Existing RFF methods remain limited by the dynamicity of wireless channels caused by mobile users. The literature lacks empirical studies in realistic public-building environments with changing transmitter-receiver geometry, strong multipath fading and high human activity. Prior experiments are often conducted over small areas ($< 100\,m^{2}$), and many do not evaluate generalization when test data are collected on different days. 

From an architectural perspective, transformer-based modeling remains underexplored in CSI-based RFF, despite CSI naturally forming a structured temporal representation suited to capturing long-range dependencies. The use of 3D-CNNs in CSI-based RFF is similarly limited, even though CSI can be represented as a time--frequency--feature tensor naturally suited to 3D convolution. 

Finally, the continued dominance of the closed-set assumption is a significant limitation; practical authentication systems must also reject previously unseen rogue devices, yet open-set classification remains far less explored in CSI-based RFF.

\textbf{\textit{Contributions:}}
To address these gaps, this work makes the following contributions:
\begin{itemize}
    \item We propose \textit{ContraCSI}, a contrastive learning framework for closed-set transmitter classification that learns a shared embedding space for CSI and device identities, improving robustness and stability over standalone classifier baselines.
    \item We extend ContraCSI to open-set authentication through a verifier--classifier framework, where a sequential detector rejects samples with previously unseen transmission patterns before classifying accepted samples among enrolled devices.
    \item We perform dynamic CSI-based RFF experiments in a large public-building environment with approximately 300 people moving naturally on average, using low-cost ESP32 modules to evaluate performance under spatial diversity, mobility, day-to-day variability, and varying SNR conditions.
    \item We further explore indoor positioning using CSI-based distance estimation and trilateration to assess the feasibility of identification followed by localization.
\end{itemize}

\section{Related Work}

Deep learning has become the dominant paradigm in RF fingerprinting. The ORACLE method~\cite{8737463} introduced controlled transmitter impairments to improve robustness under dynamic channels using an AlexNet-based 2D CNN~\cite{krizhevsky2012imagenet} on raw IQ samples. An extensive experimental study~\cite{9063411} evaluated CNN-based RFF across varying SNR ranges and device counts, although performance degraded in practical channel conditions and channel dynamicity was not clearly specified. Complex-valued CNNs combined with Variational Mode Decomposition have also been used for satellite AIS fingerprinting at low SNR~\cite{10188589}, while a triplet-loss-enhanced CNN was applied to base-station classification across WiFi, LTE, and 5G waveforms~\cite{9348261}. A broader overview of challenges and opportunities in DL-based RFF is given in~\cite{al2024radio}. In our previous works on RFF with Chaotic Antenna Array (CAA) we used 2D CNNs using VGG and ResNet backbones \cite{mcmillen2025hardware}, \cite{ranstrom2026analytical}. More recently, 3D CNNs have been explored for satellite RFF using stacked constellation images~\cite{zhu20233d} and Density Trace Plots derived from symbol transition trajectories~\cite{huang2024deep}, where 3D input volumes capture spatiotemporal fingerprint structure and outperform 2D CNN baselines. Nevertheless, most deep-learning RFF studies still rely on 2D CNNs operating on IQ samples collected with specialized SDRs or NICs, and are typically evaluated over small areas with static or near-static device placement. As a result, many works do not establish whether the learned features generalize when the transmitter--receiver geometry changes.

Transformer architectures have also recently appeared in the RFF literature. Shen et al.~\cite{shen2021radio} used a transformer for LoRa device classification, showing that sequence-oriented modeling can be effective for RF fingerprinting. Xiao et al.~\cite{xiao2025multi} proposed MPDFormer, a Multi-Periodicity Dependency Transformer that captures both short-range and long-range periodic dependencies in RF fingerprints. Hui et al.~\cite{hui2025crossattention} introduced a cross-attention transformer to reduce the impact of channel-state variation between training and testing. Herrera-Loera et al.~\cite{herrera2026transformer} used a Vision Transformer (ViT)~\cite{dosovitskiy2020image} encoder for GFSK-modulated IoT device classification and reported up to 5\% higher accuracy than CNNs in NLOS conditions while requiring fewer training epochs. Despite these promising results, existing transformer-based RFF works operate on IQ samples. To the best of our knowledge, transformer-based architectures have not been applied to CSI-based RF fingerprinting, even though CSI forms a structured temporal representation in which subcarrier responses evolve across packets and may benefit from long-range dependency modeling.

Open-set RF fingerprinting has also received increasing attention, since practical authentication systems must reject previously unseen rogue transmitters rather than only classify known devices. Karunaratne et al.~\cite{karunaratne2021openset} proposed generative outlier augmentation for open-set RF fingerprinting. Wang et al.~\cite{wang2023openset} improved prototype learning using consistency-based regularization and online label smoothing for unknown-device rejection. More recent studies have investigated multi-task prototype learning~\cite{ma2025mtplopenset} and joint prediction with Siamese comparison~\cite{cai2025jrffpsc} for open-set RFF. However, open-set classification remains much less explored than closed-set RFF, especially under realistic dynamic channel conditions. Most open-set studies use IQ samples collected with SDRs in controlled settings and do not address the challenge of separating position-dependent channel signatures from true device fingerprints when rejecting unknown transmitters.

CSI-based RFF remains comparatively underexplored. Prior work includes deep auxiliary learning that converts CSI to IQ samples using the WiSig dataset~\cite{hanna_wisig_2022} for knowledge transfer~\cite{10.1007/978-981-97-5609-4_2}, achieving 82\% accuracy on real data. CSI fingerprint distance has also been used for indoor localization~\cite{article}. Kong and Chen~\cite{10793404} proposed DeepCRF for channel-resilient WiFi device identification using CSI, reporting nearly 100\% accuracy on synthetic data at 40~dB SNR but lower performance on real data. Notably, in their real-world dataset the relative transmitter--receiver positions remain fixed across collection points, which greatly limits channel dynamicity. Tools for low-cost CSI collection on commodity hardware such as the ESP32 have been demonstrated in~\cite{Hern2006:Lightweight}. Overall, existing CSI-based RFF studies predominantly operate in static or quasi-static settings where transmitter--receiver geometry is held constant. Under such conditions, the learned representations inevitably encode the position-dependent channel factor $h_k$ together with the device-specific impairment $\psi_k$, leading to fingerprints that do not transfer well to new locations. Experiments are usually conducted over small areas ($<100\,m^2$) with few collection points, cross-day evaluation is rare, and no prior CSI-based RFF work combines a truly dynamic collection setting with mobile devices across diverse locations, transformer or 3D CNN architectures, open-set authentication, and low-cost ESP32 hardware in a realistic public-building environment with high human activity.

\section{Proposed RFF System}
\label{sec:proposed_sys}

\begin{figure}[t]
    \centering
    \includegraphics[width=\columnwidth]{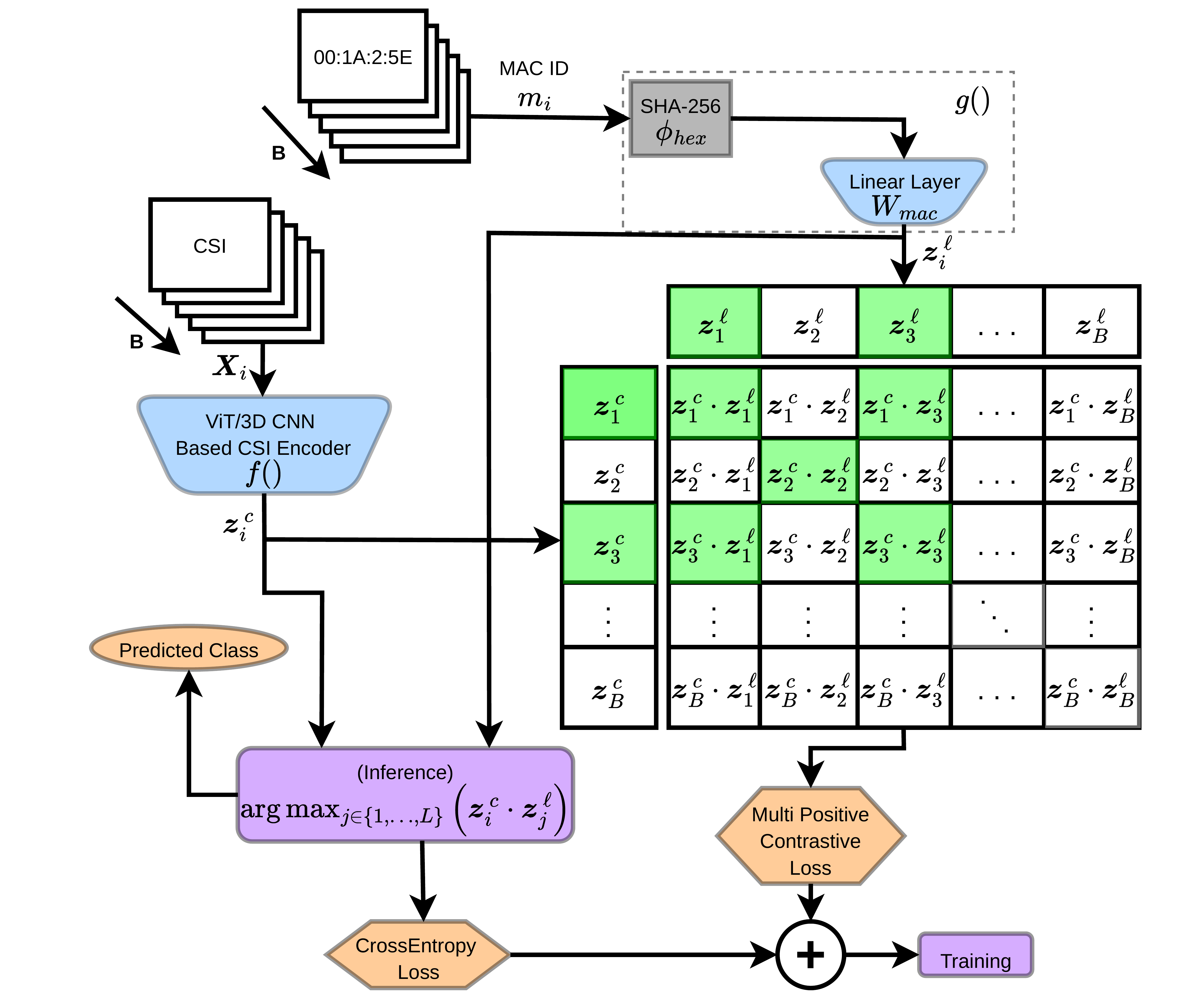}
    \caption{ContraCSI architecture. CSI windows are mapped by a ViT- or 3D CNN-based CSI encoder into CSI embeddings, while transmitter MAC-ID labels are converted by SHA-256 hashing and a learnable linear layer into label embeddings in the same space. Their pairwise similarities form a similarity matrix used to optimize a multi-positive contrastive loss. At inference, the predicted class is obtained by choosing the max-similarity class.}
    \label{fig:model_architecture}
\end{figure}

The proposed CSI-based RFF authentication system is illustrated in Fig.~\ref{fig:model_architecture}. In wireless communications, CSI represents the transfer function of the channel between transmitter and receiver, capturing multipath fading, reflection, scattering, and hardware-induced distortions. Since the same receiver is used throughout, we consider only transmitter-side distortions. The received signal on subcarrier $k$ is modeled as
\begin{equation}
    r_k = h_k \cdot \psi_k \cdot s_k + n_k,
\end{equation}
where $s_k$ is the transmitted symbol, $r_k$ is the received symbol, $h_k \in \mathbb{C}$ is the wireless channel factor, $\psi_k \in \mathbb{C}$ is the transmitter hardware impairment factor, and $n_k$ is additive white Gaussian noise. The estimated CSI for the $k^{\text{th}}$ subcarrier is
\begin{equation}
    \hat{H}_k = \frac{r_k}{s_k} = h_k \cdot \psi_k + \varepsilon_k,
    \label{eq:csi}
\end{equation}
where $\varepsilon_k = n_k/s_k$ is the normalized noise term. Thus, the device-specific distortions are embedded in CSI, but remain entangled with the position-dependent channel factor $h_k$, making robust fingerprint learning challenging in dynamic environments.

A set $\mathcal{L}$ of legitimate WiFi transmitters is registered with the authenticator. The receiver is an ESP32 SoC equipped with a lightweight CSI toolkit~\cite{Hern2006:Lightweight} that estimates CSI from each received packet. A legitimate transmitter sends a known preamble symbol vector $\vec{s}$, and the corresponding received signal vector $\vec{r}$. From the received signal, the toolkit computes the CSI estimate $\hat{\vec{H}}$ across all 64 subcarriers and forwards it to the proposed RFF system. Here, $\hat{\vec{H}} \in \mathbb{R}^{64\times 2}$ denotes the matrix formed by stacking the real and imaginary parts of the estimated CSI coefficients for all subcarriers; its explicit construction is given in Section~\ref{subsec:csi-estimation}.

Authentication is formulated as an open-set multi-class classification problem. The model outputs one of $L+1$ possible classes, where the first $L$ classes correspond to legitimate transmitters and class $L+1$ represents non-legitimate devices. Authentication succeeds when a CSI sample from legitimate transmitter $i$ is assigned to class $i$, or when a CSI sample from a non-legitimate transmitter is assigned to class $L$; any other decision is an error. For localization, CSI from at least three designated anchor transmitters is used. After CSI identification through RFF, using the known anchor coordinates $(x_{a_1},y_{a_1})$, $(x_{a_2},y_{a_2})$, $(x_{a_3},y_{a_3})$ together with the estimated distances to the anchor nodes using CSI, the system estimates the receiver position $(\hat{x}_r,\hat{y}_r)$. 

\subsection{CSI Estimation}
\label{subsec:csi-estimation}

The CSI toolkit estimates the channel using the known Long Training Field (LTF) symbols in WiFi preambles. This lightweight toolkit, originally designed for WiFi sensing, enables low-cost CSI collection on ESP32 modules~\cite{Hern2006:Lightweight}. Each of the 64 OFDM subcarriers yields a complex channel coefficient. The collected data form a 128-integer array in which each consecutive pair represents the real and imaginary parts of one subcarrier. We reshape this into
\begin{equation}
    \hat{\vec{H}} =
    \begin{bmatrix}
    \Re(\hat{H}_1) & \Im(\hat{H}_1) \\
    \Re(\hat{H}_2) & \Im(\hat{H}_2) \\
    \vdots & \vdots \\
    \Re(\hat{H}_{64}) & \Im(\hat{H}_{64})
    \end{bmatrix}
    \in \mathbb{R}^{64 \times 2},
\end{equation}
where $\Re(\hat{H}_k)$ and $\Im(\hat{H}_k)$ denote the real and imaginary parts of the CSI for the $k^{\text{th}}$ subcarrier. Each CSI sample $\hat{\vec{H}}$ corresponds to one WiFi packet. Successive packets from the same transmitter yield a time series of CSI samples. In addition to CSI, Received Signal Strength Indicator (RSSI), and noise floor are also recorded for SNR analysis while MAC addresses are collected as device labels.

\subsection{Contrastive Learning-based CSI Model}
\label{subsec:CLIP-CSI_model}

We introduce \textit{ContraCSI}, a contrastive learning framework for CSI-based transmitter identification. As shown in Fig.~\ref{fig:model_architecture}, the model jointly learns CSI and device-ID embeddings in a shared space, so that CSI embeddings move closer to the correct transmitter embedding and farther from the others. We consider two CSI encoder families: a Vision Transformer (ViT) encoder as the primary model, and 3D CNN-based encoders as structure-preserving convolutional alternatives.

The key motivation is that CSI from the same transmitter should form compact clusters despite channel variation while CSI from different transmitters should remain separable. Since the data are labeled, we adopt supervised contrastive learning~\cite{khosla2020supervised}. However, unlike single-positive CLIP-style training~\cite{radford2021learning}, our minibatches may contain multiple samples from the same transmitter. Treating only one of them as positive would incorrectly push same-class samples apart. We therefore use a \emph{multi-positive symmetric contrastive loss} that treats all same-class items in the minibatch as positives, rather than enforcing a single positive pair per anchor.

For each transmitter, CSI windows are formed by grouping the first $N$ consecutive CSI entries for the same MAC ID. Stacking $N$ such $64{\times}2$ matrices yields the $i$-th CSI window $\vec{X}_i \in \mathbb{R}^{64 \times 2 \times N}$, where $i$ indexes the training or test sample. Z-score normalization is applied before the CSI encoder by computing the mean and standard deviation from the training data.

\subsubsection{CSI Encoder}
\label{subsec:csi-encoder}

We use transformer-based and convolution-based encoders tailored to the native CSI window $\vec{X}_i \in \mathbb{R}^{64\times 2\times N}$. The CSI encoder, denoted by $f$, maps each input window $\vec{X}_i$ to a $D$-dimensional embedding vector which is subsequently $L_2$-normalized before contrastive training as shown below:
\begin{equation}
\tilde{\vec{z}}_i^c = f(\vec{X}_i), \qquad
\vec{z}_i^{\,c} = \frac{\tilde{\vec{z}}_i^{\,c}}{\left\|\tilde{\vec{z}}_i^{\,c}\right\|_2}\in \mathbb{R}^{D}.
\end{equation}

\textbf{ViT encoder.} The CSI window is rearranged from $[B,64,2,N]$ to $[B,2,64,N]$, where $B$ denotes the batch sizes, bilinearly resized to $224{\times}224$, and expanded to three channels by duplicating one of the two channels, yielding $[B,3,224,224]$. A pretrained ViT base model~\cite{vit-hface} processes this input. The CLS-token representation is projected, normalized with LayerNorm, and $L_2$-normalized to produce the embedding vector $\vec{z}_i^c$. Transformer-based modeling is attractive because it can capture long-range dependencies across packets and global interactions across subcarriers, and has shown strong performance in RF fingerprinting~\cite{al2024radio}.

\textbf{3D-CNN encoders.} The CSI window is processed as a native 3D tensor without image resizing, preserving local spatiotemporal structure across subcarriers, feature channels, and packets. We study two variants:
\begin{itemize}
    \item A lightweight custom 3D-CNN using input $[B,1,N,64,2]$, three $3{\times}3{\times}3$ convolutions with temporal and spatial downsampling, global 3D average pooling, and a linear head.
    \item A modified R3D18~\cite{tran2018closer} with the standard topology retained but the input stem adapted to two channels. The CSI window is formatted as $[B,2,N,64,1]$.
\end{itemize}
For comparison, lightweight three-layer 2D-CNN encoders are also evaluated. In all contrastive variants, the CSI encoder outputs the embedding $\vec{z}_i^c \in \mathbb{R}^{D}$, and the MAC ID labels are mapped to the same space and $L_2$-normalized, so cosine similarity equals the dot product.

\subsubsection{Label Encoder}

The label encoder, denoted by $g$, maps each class label (MAC ID) $m_i \in \{1,\dots,L\}$ to a compact unit-length $D$-dimensional embedding vector
\[
\vec{z}_i^\ell = g(m_i) \in \mathbb{R}^{D},
\]
for cosine-similarity comparison with CSI embeddings. It combines a deterministic base code for each MAC ID with a small learnable projection.

Each MAC address is converted into a fixed base vector by hashing its canonical hex string with SHA-256. The 32-byte hash is tiled or truncated to the required dimension, mapped to real values, and z-scored to form a deterministic non-learnable code. A learnable linear projector adapts these codes, and the result is $L_2$-normalized:
\begin{equation}
\tilde{\vec{z}}_i^\ell = \phi_{\mathrm{hex}}(m_i)\,W_{\mathrm{mac}}, \qquad
\vec{z}_i^\ell = \frac{\tilde{\vec{z}}_i^\ell}{\left\|\tilde{\vec{z}}_i^\ell\right\|_2},
\end{equation}
where $\phi_{\mathrm{hex}}(m_i)\in \mathbb{R}^{H}$ is the fixed code for class label $m_i$ and $W_{\mathrm{mac}}\in\mathbb{R}^{H\times D}$ is the learnable projector. Only $W_{\mathrm{mac}}$ is trained; the base codes remain fixed.

\subsubsection{Loss Function}
\label{subsec:loss-function}

Given a minibatch $\mathcal{B}=\{(\vec{X}_i,m_i)\}_{i=1}^{B}$, where $\vec{X}_i$ is the CSI window for sample $i$ and $m_i \in \{1,\dots,L\}$ is the class label corresponding to its transmitter MAC ID, we use the CSI and label embeddings defined above.

Using a trainable temperature parameter $\tau$, the scaled similarity is computed via dot product as
\[
S_{ij} = \vec{z}_i^c \cdot \vec{z}_j^\ell/\tau,
\]
and the positive set is defined as
\[
\mathcal{P}(i)=\{j \in \{1,\dots,B\}\mid m_j=m_i,\; j\neq i\}.
\]

The CSI$\to$label (row-wise) loss pushes the CSI embedding towards the label embedding of its class and away from the label embeddings of other classes:
\begin{equation}
\mathcal{L}_{c\to \ell}
=
\frac{1}{B}\sum_{i=1}^{B}
\left[
-\log
\frac{
\displaystyle\sum_{j\in\mathcal{P}(i)} \exp(S_{ij})
}{
\displaystyle\sum_{\substack{k=1\\k\neq i}}^{B}\exp(S_{ik})
}
\right].
\label{eq:mp-cl-3p}
\end{equation}

Similarly, the label$\to$CSI (column-wise) loss pushes the label embedding towards the CSI embeddings of samples from that class and away from the CSI embeddings of samples from other classes:
\begin{equation}
\mathcal{L}_{\ell\to c}
=
\frac{1}{B}\sum_{j=1}^{B}
\left[
-\log
\frac{
\displaystyle\sum_{i\in\mathcal{P}(j)} \exp(S_{ij})
}{
\displaystyle\sum_{\substack{k=1\\k\neq j}}^{B}\exp(S_{kj})
}
\right].
\label{eq:mp-lc-3p}
\end{equation}
The multi-positive supervised contrastive loss is then given by
\begin{equation}
\mathcal{L}_{\mathrm{con}}
= \tfrac{1}{2}\left(\mathcal{L}_{c\to \ell}+\mathcal{L}_{\ell\to c}\right).
\end{equation}
Compared with the single-positive CLIP loss, this formulation leverages all same-class items in the batch as positives by pooling their similarity mass, thereby stabilizing training and reducing false negatives.

\textit{Cross-entropy loss (optional).}
Rather than using an explicit classifier head, we predict the class label for instance $i$ as 
\[
j^*=\arg\max_{j \in \{1,\dots,L\}} S_{ij}
\]
The cross-entropy loss
\begin{equation}
\mathcal{L}_{\mathrm{CE}}
= \frac{1}{B}\sum_{i=1}^{B}
\left[-\log\frac{\exp(S_{ij^*})}{\sum_{k=1}^{L}\exp(S_{ik})}\right].
\end{equation}
The total training loss becomes
\begin{equation}
\mathcal{L}_{\mathrm{total}}
= \mathcal{L}_{\mathrm{con}}
+ \alpha\,\mathcal{L}_{\mathrm{CE}},
\qquad \alpha>0,
\label{eq:alpha}
\end{equation}
where $\alpha$ is selected using validation set. We optimize the parameters of $f$ and $g$ with AdamW, and $\tau$ is learned and clamped (e.g., $\tau^{-1}\le 100$).

\subsection{CUSUM-Based Open-Set Authentication}
\label{subsec:cusum-openset}

\begin{figure}[t]
    \centering
    \includegraphics[width=\columnwidth]{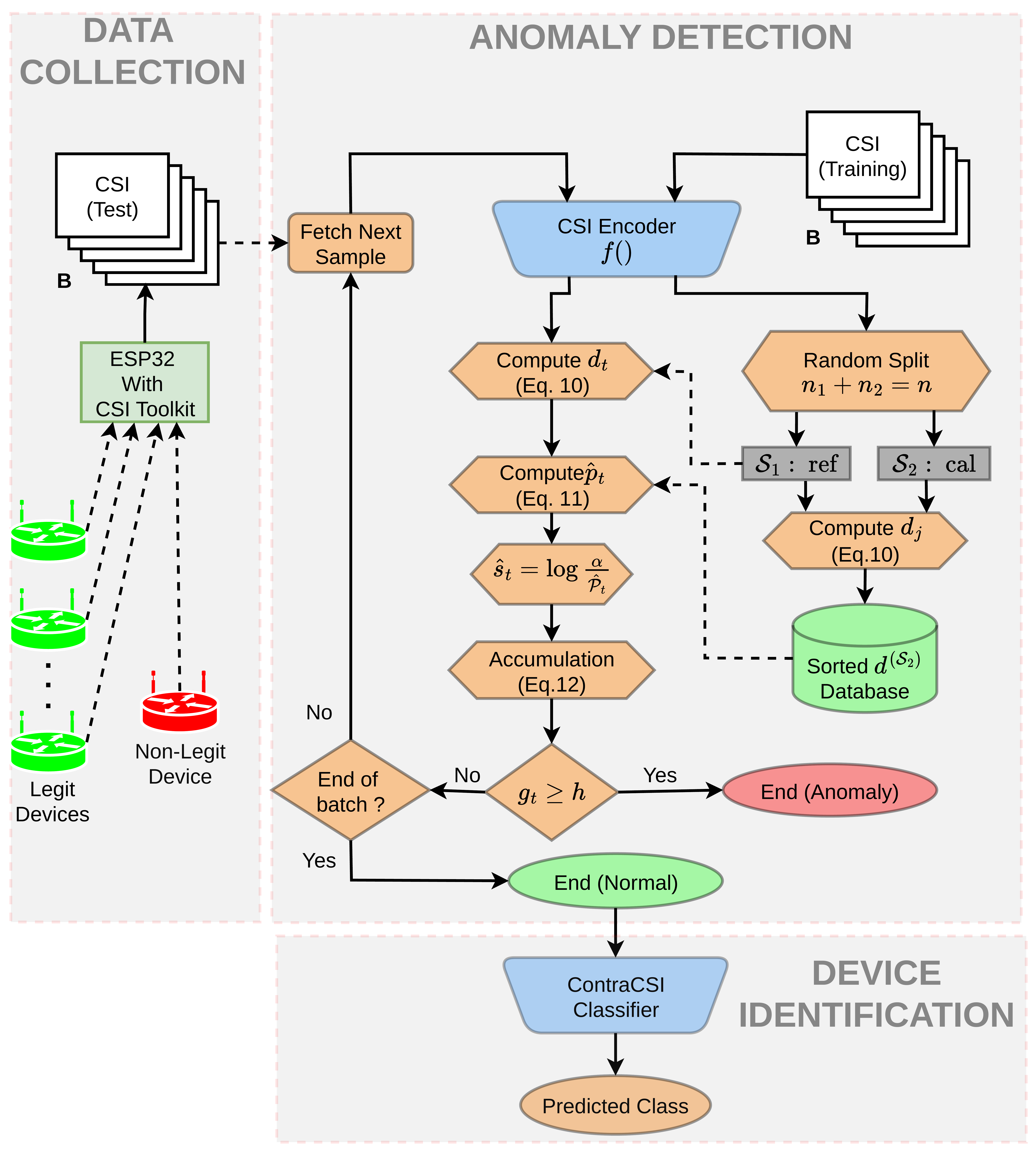}
    \caption{Open-set authentication pipeline of the proposed ContraCSI framework. CSI test windows are first mapped to embeddings by the Lite3D-CNN-ContraCSI encoder and evaluated by a GEM--CUSUM verifier. Accepted samples are passed to the ContraCSI classifier for closed-set device identification, while anomalous samples are rejected as non-legitimate devices.}
    \label{fig:openset_architecture}
\end{figure}

In addition to closed-set transmitter identification, we extend ContraCSI to an \emph{open-set} setting in which previously unseen transmitters must be rejected rather than forced into one of the known classes. As illustrated in Fig.~\ref{fig:openset_architecture}, we use a verifier--classifier design: a GEM--CUSUM anomaly detector~\cite{kurt2020real} first tests whether an incoming CSI sequence is consistent with the distribution of legitimate devices, and only accepted samples are forwarded to the closed-set ContraCSI classifier for device-ID prediction.

The verifier operates on the CSI embedding space learned by ContraCSI. Using only legitimate training data, the encoder produces nominal embeddings, which are randomly partitioned into a reference set $\mathcal{S}_1$ with $n_1$ samples and a calibration set $\mathcal{S}_2$ with $n_2$ samples. The CSI embeddings $\{\vec{z}_i^c = f(\vec{X}_i)\}_i$ are grouped into non-overlapping batches of length $T$ for sequential GEM--CUSUM testing.

For each CSI embedding in the calibration set $\vec{z}_j^c \in \mathcal{S}_2$, let
$\vec{z}_{(i)}^{\mathcal{S}_1}(\vec{z}_j^c)\in\mathcal{S}_1$
denote the $i$-th nearest neighbor of $\vec{z}_j^c$ in $\mathcal{S}_1$. The  $k$-nearest-neighbor ($k$NN) distance statistic 
\begin{equation}
d_j^{(\mathcal{S}_2)}=\sum_{i=1}^{k}\left\|\vec{z}_j^c-\vec{z}_{(i)}^{\mathcal{S}_1}(\vec{z}_j^c)\right\|_2
\end{equation}
are used to empirically evaluate the anomaly score of test instances.

At test time, each incoming CSI window $\vec{X}_t$ is mapped to an embedding $\vec{z}_t^c = f(\vec{X}_t)$, and its distance statistic $d_t$ is computed relative to $\mathcal{S}_1$:
\begin{equation}
d_t=\sum_{i=1}^{k}\left\|\vec{z}_t^c-\vec{z}_{(i)}^{\mathcal{S}_1}(\vec{z}_t^c)\right\|_2.
\end{equation}
An empirical $p$-value is then obtained by comparing $d_t$ against the calibration distances from $\mathcal{S}_2$:
\begin{equation}
\hat{p}_t = \frac{1}{n_2}\sum_{j=1}^{n_2} \vec{1}\{ d_j^{(\mathcal{S}_2)} \ge d_t \}.
\end{equation}
Following GEM--CUSUM, we compute
\begin{equation}
\hat{s}_t=\log\frac{\alpha}{\hat{p}_t+\epsilon},
\qquad
g_t=\max\!\left(0,\,g_{t-1}+\hat{s}_t\right),
\end{equation}
where $\alpha<\exp(-1)$ is the statistical significance level optimized using the validation set, $\hat{p}_t$ is the empirical tail probability under the nominal calibration set, $\epsilon$ is a small constant used to avoid division by zero, and $g_t$ is the accumulated anomaly statistic. An alarm is raised once $g_t \ge h$, where $h$ is a user-defined threshold, controlling the trade-off between false and true positive rates. This sequential accumulation is valuable in dynamic indoor settings because weak anomaly evidence across multiple CSI windows can be integrated over time for better false alarm control.

If the CUSUM statistic does not exceed the threshold at any step $t \in \{1,\dots,T\}$ within a batch, the batch is accepted as legitimate and forwarded to the closed-set ContraCSI classifier for device-ID prediction. If an alarm is triggered at any step, the batch is rejected as belonging to an unseen or non-legitimate transmitter. Thus, the proposed framework supports open-set CSI identification within a single authentication pipeline.

\subsection{CSI-Based Trilateration}\label{sec:trilateration}

We implement a CSI-based distance regression and trilateration framework for indoor localization. After transmitter identification through RF fingerprinting, a fully connected network predicts the distance between the receiver and the transmitter, which is assumed to be anchored at a fixed location, from raw CSI data. Each $64 \times 2$ CSI input is flattened and passed through five linear layers (128$\to$256$\to$128$\to$64$\to$32$\to$1) with ReLU activations and Dropout(0.2) on the first two layers. The dataset is constructed from CSI logs where each row is filtered by MAC address and reshaped into a $(64,2)$ tensor, labeled with the Euclidean ground-truth distance $d_i = \sqrt{(x_a^i - x_r)^2 + (y_a^i - y_r)^2}$, where $(x_a^i,y_a^i)$ and $(x_r,y_r)$ are the coordinates of the $i$-th anchor and the receiver. The objective is to estimate the distances $\{d_i\}$ to at least three anchors and in turn $(x_r,y_r)$ through trilateration.

In the experiments, three separate models are trained, one per anchor. The predicted distances $\hat{d}_1, \hat{d}_2, \hat{d}_3$ and the known anchor coordinates $(x_a^i, y_a^i),~i=1,2,3$, are used to solve linearized pairwise difference-of-distance equations for 2D position estimation \cite{}. Localization performance is evaluated using the mean-squared-error metric:
\[
\text{MSE} = \frac{1}{N} \sum_{j=1}^{N} \left[(x_r^j - \hat{x}_r^j)^2 + (y_r^j - \hat{y}_r^j)^2\right],
\]
where $N$ denotes the number of receiver location instances in the dataset. 

\section{Results and Discussion}

\begin{figure}[t]
    \centering
    \includegraphics[width=\columnwidth]{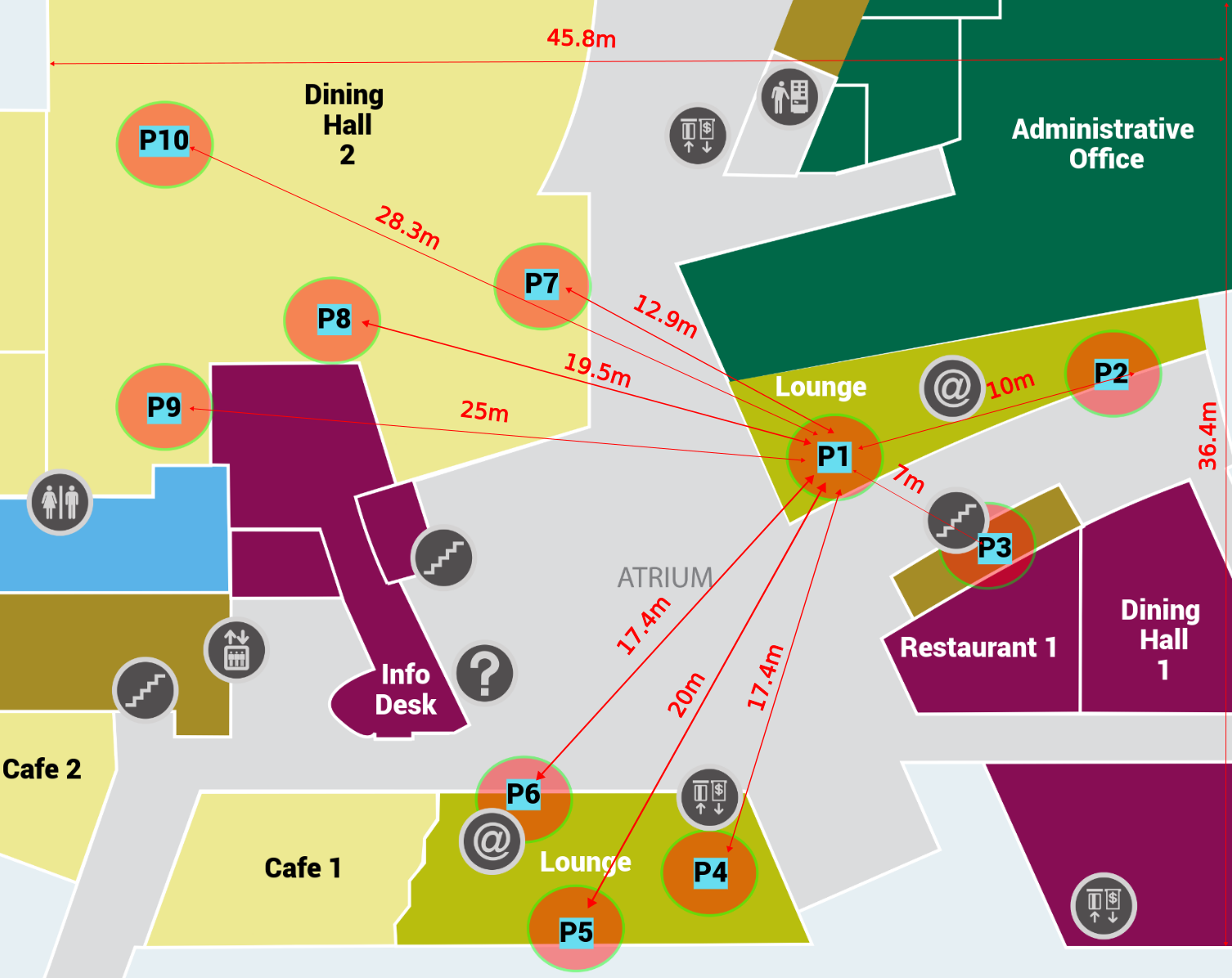}
    \caption{Floor map of the Marshall Student Center showing the 10 CSI data collection positions (P1--P10) used for dynamic RF fingerprinting experiments. The measurement locations span three zones with different levels of human activity and multipath conditions, covering approximately $1700~m^2$.}
    \label{fig:msc_map}
\end{figure}

We evaluate the proposed framework in three settings following the practical authentication flow: open-set authentication with unseen transmitters, closed-set identification of accepted legitimate transmitters, and indoor localization. The open-set and closed-set identification experiments were conducted in a university student center during active hours with hundreds of people moving in the environment, while the localization experiments were conducted in a convention hall with high human activity.

\subsection{Open-Set Authentication}

\begin{table*}[!t]
\centering
\caption{Open-set GEM--CUSUM performance across encoder backbones for different numbers of unseen MACs. Each entry reports F1/AUC.}
\label{tab:openset_all_models}
\scriptsize
\setlength{\tabcolsep}{3.5pt}
\renewcommand{\arraystretch}{1.05}
\resizebox{0.95\textwidth}{!}{%
\begin{tabular}{|c|c|c|c|c|c|c|}
\hline
\textbf{\# Unseen MACs} 
& \textbf{Lite3D-CNN} 
& \textbf{Lite3D-CNN-Contra} 
& \textbf{R3D18} 
& \textbf{R3D18-Contra} 
& \textbf{ViT-CE} 
& \textbf{ViT-Contra} \\
\hline
1 
& 0.87 / 0.9295 
& \textbf{0.97 / 0.9611} 
& 0.95 / 0.9755 
& \textbf{0.97 / 0.9895} 
& 0.96 / 0.9824 
& 0.92 / 0.9585 \\
\hline
2 
& 0.84 / 0.9017 
& \textbf{0.99 / 0.9994} 
& 0.82 / 0.7394 
& 0.93 / 0.9842 
& 0.88 / 0.9008 
& 0.72 / 0.7155 \\
\hline
3 
& \textbf{1.00 / 0.9999} 
& \textbf{1.00 / 1.0000} 
& \textbf{1.00 / 1.0000} 
& \textbf{1.00 / 1.0000} 
& \textbf{1.00 / 0.9999} 
& \textbf{1.00 / 1.0000} \\
\hline
4 
& 0.91 / 0.9596 
& \textbf{0.96 / 0.9744} 
& 0.88 / 0.9544 
& 0.85 / 0.9281 
& 0.91 / 0.9000 
& 0.85 / 0.9168 \\
\hline
5 
& 0.94 / 0.9575 
& \textbf{0.96 / 0.9694} 
& 0.92 / 0.9652 
& 0.93 / 0.9785 
& 0.94 / 0.9489 
& 0.93 / 0.9699 \\
\hline
\textbf{Average} 
& 0.91 / 0.9496 
& \textbf{0.98 / 0.9800} 
& 0.91 / 0.9269 
& 0.94 / 0.9761 
& 0.94 / 0.9464 
& 0.86 / 0.9125 \\
\hline
\end{tabular}%
}
\end{table*}


\begin{figure}[t]
    \centering
    \includegraphics[width=\columnwidth]{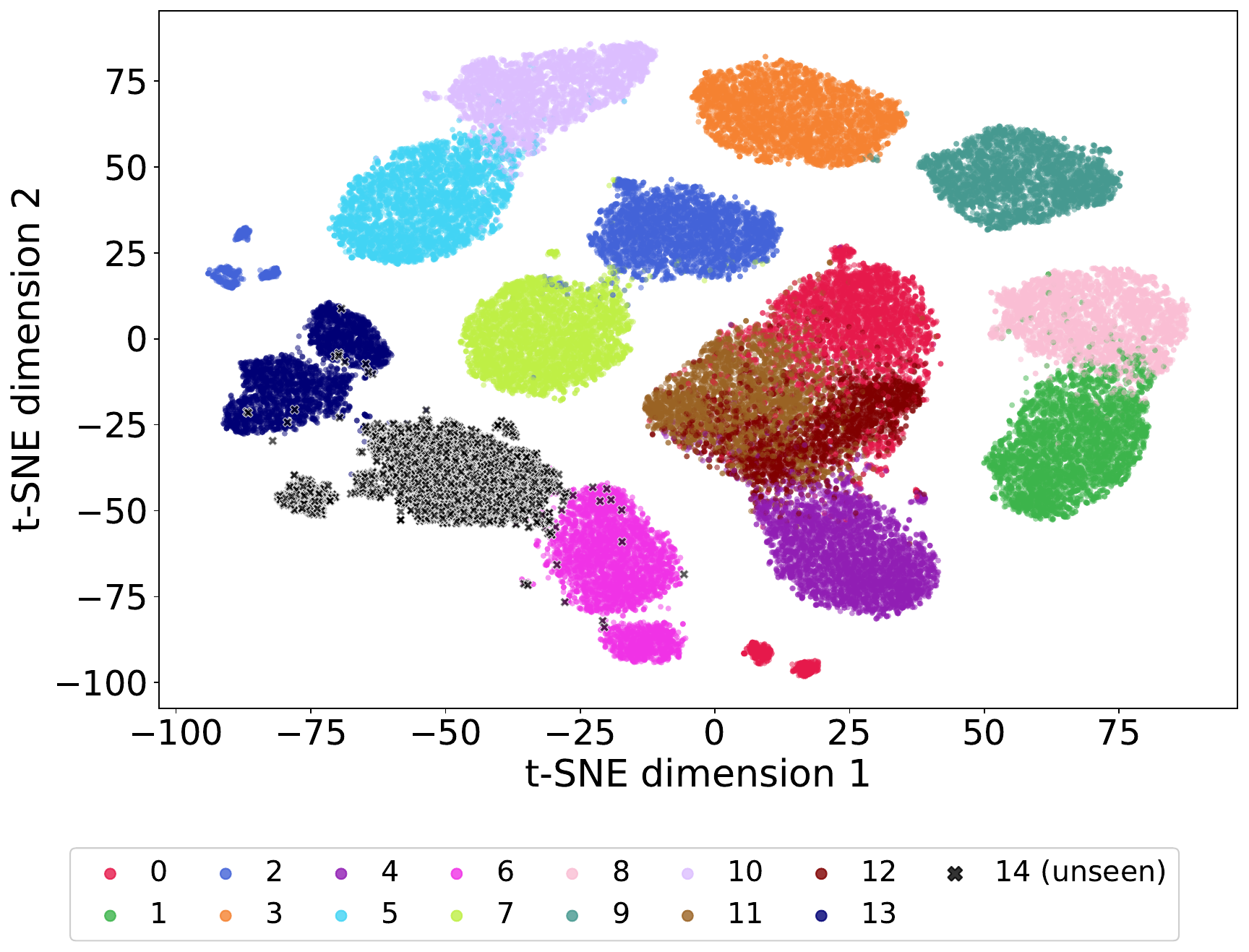}
    \caption{Density-based t-SNE visualization of the full test split for the single-anomaly experiment at $N=4$. The unseen transmitter remains separated from most legitimate classes in the held-out test embedding space.}
    \label{fig:tsne_test_openset}
\end{figure}

\begin{figure}[t]
    \centering
    \includegraphics[width=\columnwidth]{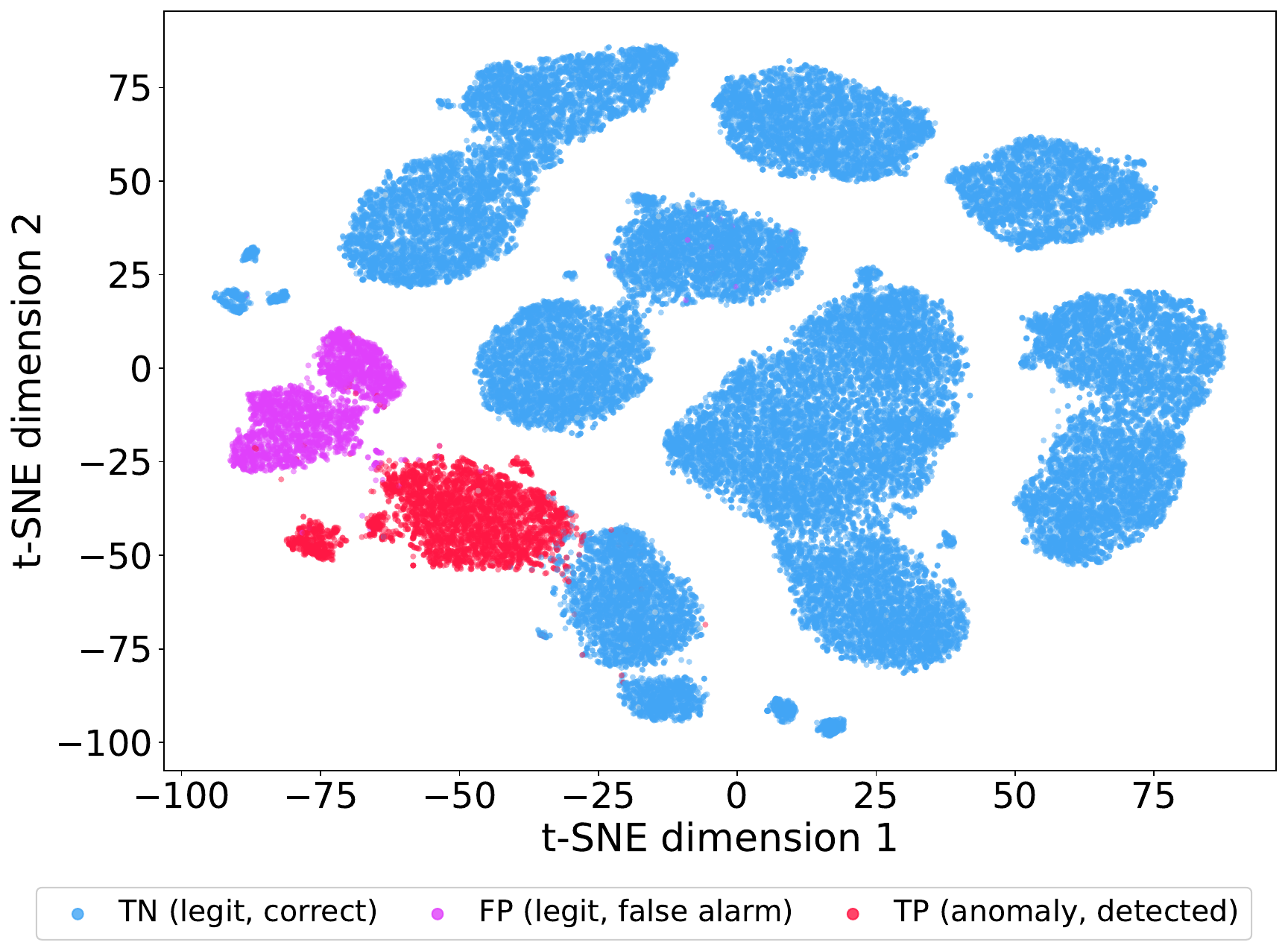}
    \caption{Density-based t-SNE visualization of GEM--CUSUM detection outcomes for the single-anomaly experiment at the selected operating threshold $h=26.26$. True negatives correspond to legitimate samples correctly accepted, false positives correspond to legitimate samples incorrectly rejected, and true positives correspond to anomaly samples correctly detected.}
    \label{fig:tsne_detection_outcome}
\end{figure}

The open-set formulation itself was introduced earlier; here we focus on the experimental setup, model comparison, and final verifier selection. In the proposed authentication flow, open-set verification is performed before closed-set classification. The receiver first determines whether an incoming CSI sequence is consistent with the enrolled legitimate transmitters. Only samples accepted as legitimate are then forwarded to the closed-set classifier for device-ID prediction. This ordering reflects the practical requirement that an authentication system must first reject unseen or non-legitimate devices rather than forcing every received signal into one of the known transmitter classes.

The open-set and closed-set experiments use the same dynamic CSI dataset collected at 10 different positions on the first floor of the Marshall Student Center at the Tampa campus of University of South Florida, denoted by P1 to P10 in Fig.~\ref{fig:msc_map}, covering approximately $1700\,\mathrm{m}^2$. The area was divided into three zones to capture different levels of human activity and multipath effects. Zone 1 (P1--P3) had fewer reflecting surfaces, Zone 2 (P4--P6) had the highest human activity, and Zone 3 (P7--P10) had both high human activity and stronger multipath. This varying transmitter-receiver geometry ensured that the identification performance is attributed to the RFF signatures, not the position signatures present in static data collection. Data were collected from March 2025 to May 2025 using a NodeMCU ESP-WROOM-32 and the ESP32 CSI-Toolkit \cite{Hern2006:Lightweight} in Active CSI collection mode. Each sample contains the 128-integer CSI vector, MAC ID, RSSI, and noise floor information.

For a longitudinal study that evaluates the open-set authentication and closed-set identification performance over various channel conditions, we used CSI data from 15 permanent MAC IDs, which belong to wireless routers, across P1 to P10, filtering out the temporary personal devices like cell phones and laptops. The average SNR, computed as
\begin{equation}
\mathrm{SNR}_{\mathrm{dB}} = \mathrm{RSSI}_{\mathrm{dBm}} - \mathrm{NoiseFloor}_{\mathrm{dBm}},
\end{equation}
spanned approximately $12$ to $29\,\mathrm{dB}$, indicating realistic low-, medium-, and high-SNR operating conditions.

For open-set authentication, we evaluate GEM--CUSUM on the embedding spaces produced by multiple encoder backbones, including Lite3D-CNN, R3D18, ViT, and their corresponding contrastive variants. Each encoder is trained using only legitimate transmitters, while unseen transmitters are completely excluded during training and introduced only during testing. CSI packets are grouped into non-overlapping windows of $N=4$ packets, mapped to embeddings, and processed by GEM--CUSUM over non-overlapping trials of length $T=20$. The legitimate embeddings are split into reference and calibration sets, and the GEM statistic uses $k=10$ nearest neighbors. To evaluate robustness under different open-set conditions, the number of unseen MACs is varied from 1 to 5.

Table~\ref{tab:openset_all_models} summarizes the open-set GEM--CUSUM performance using F1 score and ROC-AUC across all evaluated encoder backbones. F1 measures the operating-point balance between precision and recall, while AUC measures threshold-independent separation between legitimate and unseen transmitters. Thus, the table compactly reports both the selected-threshold detection performance and the overall ranking quality of the anomaly scores.

Lite3D-CNN-Contra gives the strongest overall open-set performance in terms of F1. Compared with the non-contrastive Lite3D-CNN, it improves the average F1 score from 0.91 to 0.98. It also remains stable across all unseen-transmitter settings, achieving F1 scores of 0.97, 0.99, 1.00, 0.96, and 0.96 for 1, 2, 3, 4, and 5 unseen MACs, respectively. Its corresponding AUC values are 0.9611, 0.9994, 1.0000, 0.9744, and 0.9694, with an average AUC of 0.9800. These results show that the contrastive Lite3D-CNN embedding provides a strong nominal manifold for GEM--CUSUM and enables reliable rejection of unseen transmitters.

The R3D18 results provide additional evidence that contrastive learning can improve open-set authentication beyond the Lite3D-CNN case alone. R3D18-Contra improves the average F1 score from 0.91 to 0.94 compared with the standard R3D18 model. It also improves the average AUC from 0.9269 to 0.9761, indicating stronger threshold-independent separation between legitimate and unseen transmitters. Although the improvement is not uniform for every unseen-MAC setting, the overall trend supports the conclusion that contrastive learning can improve the suitability of 3D CNN-based embeddings for GEM--CUSUM-based open-set authentication.

The ViT results show a different behavior. While ViT-based models achieve the strongest closed-set identification performance, ViT-Contra does not provide the best open-set verification performance in this experiment. Its average F1 score is 0.86, compared with 0.94 for ViT-CE, and its average AUC is 0.9125, compared with 0.9464 for ViT-CE. This suggests that the best closed-set classifier is not necessarily the best open-set verifier. Closed-set classification rewards discriminative boundaries among known transmitters, whereas open-set authentication requires a stable embedding geometry that separates legitimate and unseen distributions. Therefore, the open-set model should be selected based on verifier performance, not only closed-set classification accuracy.

Based on this full comparison, Lite3D-CNN-Contra is selected as the final open-set verifier. This choice is based on its overall F1 and AUC performance across all evaluated models, not only on a pairwise comparison with Lite3D-CNN. As shown in Table~\ref{tab:openset_all_models}, Lite3D-CNN-Contra achieves the highest average F1 and the highest average AUC among the evaluated models. The results therefore support the broader conclusion that contrastive CSI embedding learning is useful for open-set authentication, particularly when combined with 3D CNN encoders that preserve the native spatiotemporal structure of CSI.

For the selected Lite3D-CNN-Contra verifier, the best case occurs with three unseen transmitters, where the verifier achieves both an F1 score of 1.00 and an AUC of 1.0000. Even with five unseen transmitters, the verifier maintains an F1 score of 0.96 and an AUC of 0.9694. Overall, these results show that GEM--CUSUM can effectively exploit the Lite3D-CNN-Contra embedding space to reject non-legitimate transmitters before closed-set device identification is applied.

For the single-anomaly case, Fig.~\ref{fig:tsne_test_openset} shows that the unseen transmitter occupies a distinct region in the learned embedding space relative to most legitimate devices in the full held-out test distribution. The corresponding detection-outcome visualization in Fig.~\ref{fig:tsne_detection_outcome} shows that true positives concentrate in the anomaly region, while false positives occur only in localized parts of the legitimate manifold, indicating that false alarms are not uniformly distributed. The AUC values reported in Table~\ref{tab:openset_all_models} further confirm that the selected verifier maintains strong open-set detection capability across different anomaly-set sizes.

\subsection{Closed-Set Identification}

\begin{figure}[t]
    \centering
    \includegraphics[width=\columnwidth]{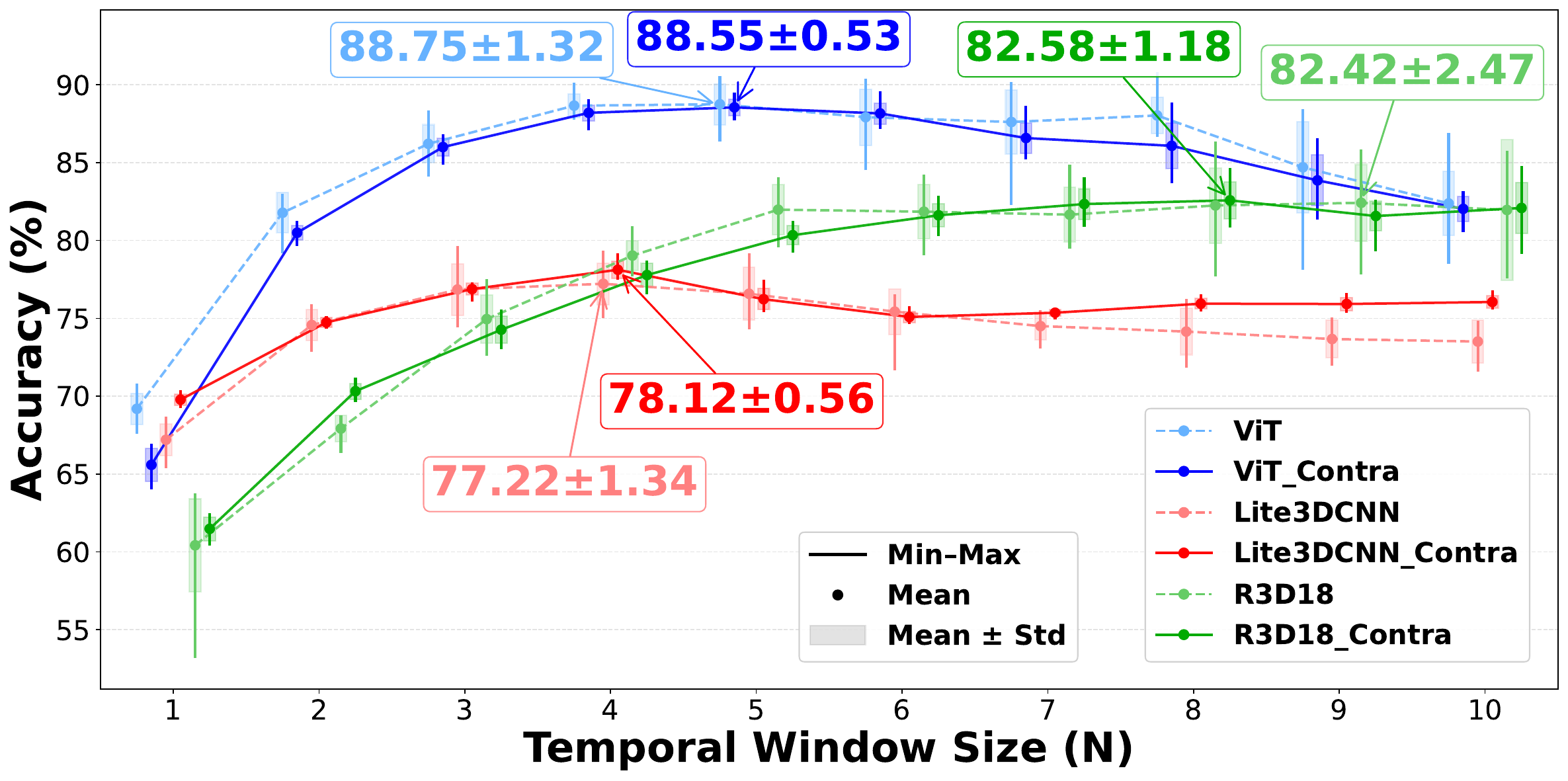}
    \caption{Top-1 test accuracy (\%) versus temporal window length $N$ for ViT, Lite3D-CNN, and R3D18, along with their ContraCSI variants. For each $N$, the marker denotes the mean across 10 seeds, the shaded vertical rectangle represents $\pm 1$ standard deviation around the mean, and the thin vertical line spans the minimum to maximum across seeds.}
    \label{fig:top1_vs_N_all_models}
\end{figure}


\begin{table*}[!t]
\centering
\caption{Macro F1 score (mean $\pm$ std) across temporal window sizes $N=1$ to $10$ for all models. The best mean F1 for each model is highlighted in bold.}
\label{tab:f1_all_models_full}
\small
\setlength{\tabcolsep}{3.5pt}
\renewcommand{\arraystretch}{0.90}
\resizebox{0.95\textwidth}{!}{%
\begin{tabular}{|c|c|c|c|c|c|c|}
\hline
\textbf{$N$} & \textbf{ViT} & \textbf{ViT-Contra} & \textbf{Lite3D-CNN} & \textbf{Lite3D-CNN-Contra} & \textbf{R3D18} & \textbf{R3D18-Contra} \\
\hline
1  & 0.6841 $\pm$ 0.0112 & 0.6468 $\pm$ 0.0120 & 0.6746 $\pm$ 0.0109 & 0.7036 $\pm$ 0.0044 & 0.6004 $\pm$ 0.0295 & 0.6098 $\pm$ 0.0074 \\
\hline
2  & 0.8101 $\pm$ 0.0138 & 0.7959 $\pm$ 0.0050 & 0.7535 $\pm$ 0.0093 & 0.7594 $\pm$ 0.0031 & 0.6762 $\pm$ 0.0078 & 0.7000 $\pm$ 0.0047 \\
\hline
3  & 0.8549 $\pm$ 0.0133 & 0.8523 $\pm$ 0.0059 & 0.7806 $\pm$ 0.0136 & 0.7856 $\pm$ 0.0030 & 0.7473 $\pm$ 0.0145 & 0.7418 $\pm$ 0.0083 \\
\hline
4  & \textbf{0.8795 $\pm$ 0.0080} & 0.8743 $\pm$ 0.0054 & 0.7892 $\pm$ 0.0090 & \textbf{0.7995 $\pm$ 0.0035} & 0.7913 $\pm$ 0.0108 & 0.7797 $\pm$ 0.0079 \\
\hline
5  & 0.8792 $\pm$ 0.0138 & \textbf{0.8769 $\pm$ 0.0058} & \textbf{0.7893 $\pm$ 0.0127} & 0.7885 $\pm$ 0.0052 & 0.8220 $\pm$ 0.0136 & 0.8082 $\pm$ 0.0053 \\
\hline
6  & 0.8686 $\pm$ 0.0205 & 0.8718 $\pm$ 0.0076 & 0.7817 $\pm$ 0.0095 & 0.7826 $\pm$ 0.0032 & 0.8250 $\pm$ 0.0169 & 0.8256 $\pm$ 0.0059 \\
\hline
7  & 0.8661 $\pm$ 0.0214 & 0.8543 $\pm$ 0.0108 & 0.7725 $\pm$ 0.0075 & 0.7854 $\pm$ 0.0019 & 0.8266 $\pm$ 0.0154 & 0.8389 $\pm$ 0.0080 \\
\hline
8  & 0.8694 $\pm$ 0.0129 & 0.8490 $\pm$ 0.0164 & 0.7687 $\pm$ 0.0101 & 0.7898 $\pm$ 0.0032 & 0.8382 $\pm$ 0.0212 & 0.8452 $\pm$ 0.0086 \\
\hline
9  & 0.8327 $\pm$ 0.0328 & 0.8236 $\pm$ 0.0188 & 0.7623 $\pm$ 0.0107 & 0.7896 $\pm$ 0.0034 & \textbf{0.8421 $\pm$ 0.0215} & 0.8419 $\pm$ 0.0065 \\
\hline
10 & 0.8072 $\pm$ 0.0237 & 0.8026 $\pm$ 0.0096 & 0.7614 $\pm$ 0.0097 & 0.7921 $\pm$ 0.0035 & 0.8403 $\pm$ 0.0416 & \textbf{0.8476 $\pm$ 0.0112} \\
\hline
\end{tabular}%
}
\end{table*}

The closed-set identification experiments use the same dynamic MSC dataset, described in the open-set authentication section. To evaluate under unseen channel conditions, we split the data temporally into training, validation, and test: for each position, all data except the last two days were used for training, the second last day for validation, and the last day for testing. Balanced subsets ensured that each MAC ID contributed similar number of samples, with approximately 70{,}000 CSI samples per device in training. All models were implemented in PyTorch and trained on a workstation with an AMD Ryzen 9 7950X and an NVIDIA RTX 4090 GPU with 24\,GB VRAM. In this stage, only samples accepted as legitimate by the verifier are classified among the enrolled transmitter identities.

We evaluated eight models: {ViT}, {ViT-Contra}, {Lite3D-CNN}, {Lite3D-CNN-Contra}, {R3D18}, {R3D18-Contra}, {2D-CNN}, and {2D-CNN-Contra}. Each experiment was repeated over 10 random seeds to evaluate robustness. For the contrastive models, we also examined the effect of the optional cross-entropy term in Eq. \eqref{eq:alpha}. For Lite3D-CNN-Contra and R3D18-Contra, including this term did not improve validation or test performance, so these models were trained using only the multi-positive contrastive objective. In contrast, for ViT-Contra, validation experiments showed that the best performance was obtained with $\alpha=0.5$, and this value was used in all reported ViT-Contra results. For the ViT, Lite3D-CNN, and R3D18 families, the temporal window length was swept from $N=1$ to $N=10$. 

Figure~\ref{fig:top1_vs_N_all_models} and Table~\ref{tab:f1_all_models_full} summarize test accuracy and macro F1, respectively. The best accuracies are 88.75$\pm$1.32 for ViT, 88.55$\pm$0.53 for ViT-Contra, 77.22$\pm$1.34 for Lite3D-CNN, 78.12$\pm$0.56 for Lite3D-CNN-Contra, 82.42$\pm$2.47 for R3D18, and 82.58$\pm$1.18 for R3D18-ContraCSI. The corresponding best macro F1 values are 0.8795$\pm$0.0080, 0.8769$\pm$0.0058, 0.7893$\pm$0.0127, 0.7995$\pm$0.0035, 0.8421$\pm$0.0215, and 0.8476$\pm$0.0112.

These observations admit a simple physical interpretation. A single CSI snapshot mixes the transmitter's hardware-imposed distortion with instantaneous channel and noise; stacking $N$ consecutive packets from the same device effectively exposes a short temporal slice of the channel response. As $N$ grows, the encoder can average out fast fading and packet-level noise while still capturing stable structure---subcarrier correlations, amplitude/phase patterns, and hardware-specific biases---that recur across packets. That is why accuracy rises sharply from $N{=}1$ to moderate $N$: the model is no longer forced to guess identity from one noisy draw, but from a small trajectory that is more informative and less brittle. The eventual plateau reflects diminishing returns: successive packets are not independent, so additional context becomes redundant once the window is long enough to span the relevant temporal scale of variation without simply over-smoothing or matching spurious short-term fluctuations.

The contrastive ContraCSI variants deserve emphasis in this setting. Supervised contrastive training explicitly aligns each CSI embedding with its device identity in a shared metric space while pushing embeddings from different devices apart. In a crowded, multipath-rich venue, class boundaries are easily blurred by zone-to-zone differences in human activity and reflection; a purely discriminative head can still memorize position- or condition-specific cues. The symmetric CSI--label objective in ContraCSI instead encourages \emph{identity-consistent} geometry: legitimate devices form tight, well-separated clusters in embedding space, which is exactly the structure one wants when authentication must survive unseen positions and temporal splits. Empirically, this shows up as not only competitive or improved peak accuracy---for example, Lite3D-CNN-ContraCSI and R3D18-ContraCSI edging their non-contrastive counterparts---but also reduced variance across random seeds, indicating more stable optimization and less sensitivity to initialization in a challenging real-world dataset. The ViT family remains strongest overall, likely because self-attention can relate subcarriers and time steps globally; yet the consistent lift from ContraCSI on the 3D CNN backbones underscores that the training objective, not only capacity, drives robustness. Together with the large gap versus flattened 2D CNN baselines (67.33$\pm$0.81 for 2DCNN and 68.92$\pm$0.36 for 2DCNN-Contra), these results support the conclusion that preserving spatiotemporal CSI structure \emph{and} learning embeddings under a contrastive identity loss are both important for dependable authentication in dynamic indoor environments.

\subsection{Trilateration Using CSI Data}

\begin{table}[!t]
\centering
\small
\caption{Root Mean Square Error (RMSE) in meters for individual anchors and trilateration-based localization.}
\label{table:rmse-trilateration}
\begin{tabular}{|c|c|c|c|c|}
\hline
\textbf{Metric} & \textbf{A1} & \textbf{A2} & \textbf{A3} & \textbf{Trilateration} \\
\hline
RMSE (m) & 0.696 & 1.029 & 0.847 & 1.180 \\
\hline
\end{tabular}
\end{table}

\begin{figure}[t]
    \centering
    \includegraphics[width=\columnwidth]{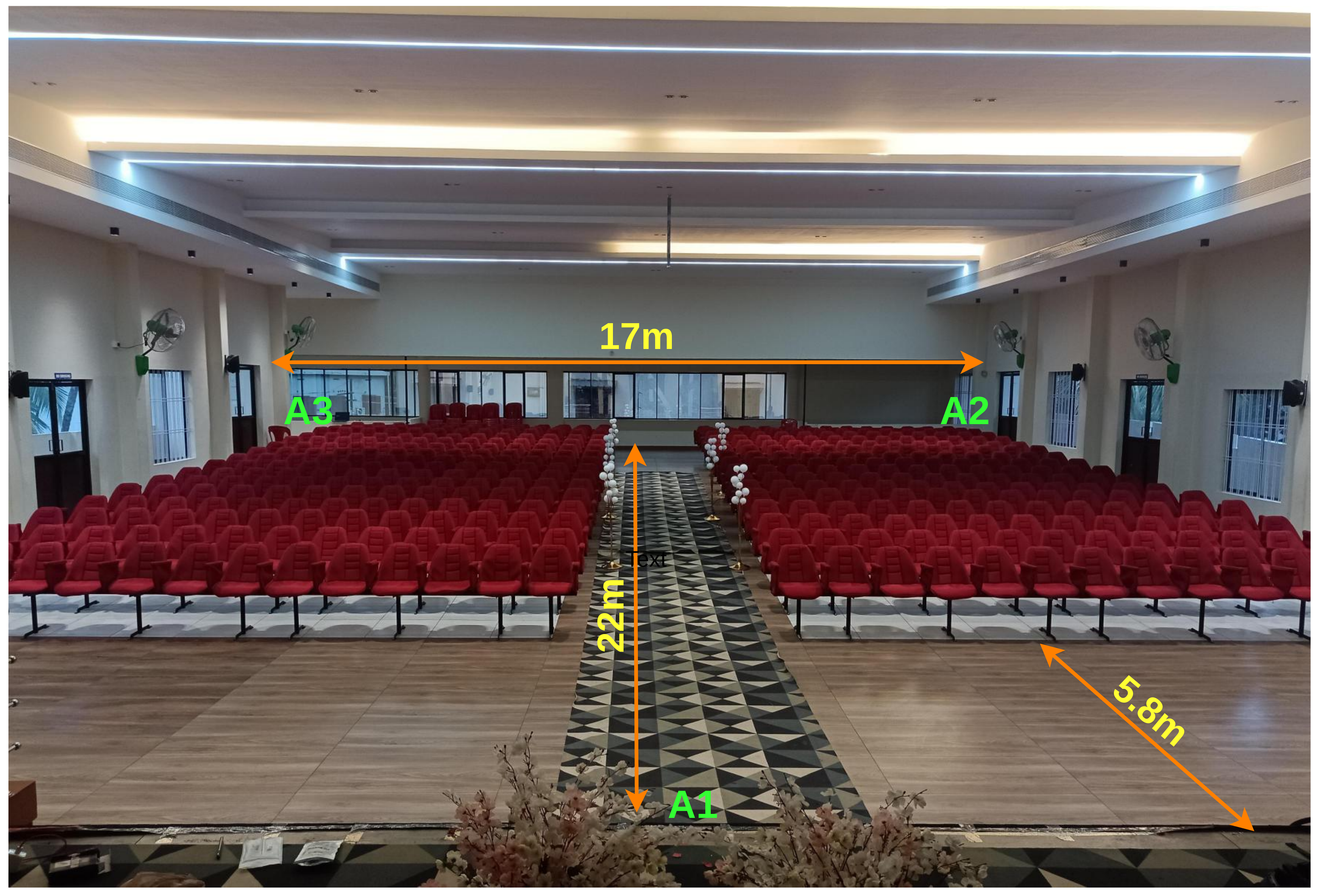}
    \caption{Trilateration map of the conference hall showing the localization region and anchor placement.}
    \label{fig:trilateration_map}
\end{figure}

As a further extension of CSI-based RFF, we explore indoor localization after authentication. Unlike authentication, the goal here is to estimate the position of a mobile receiver using signals from fixed anchor transmitters. We did not use the Marshall Student Center for the localization experiments, even though it was used for device classification and open-set authentication, because the present study considers 2D trilateration rather than 3D trilateration. Extending the setup to 3D localization would require a more complex deployment and additional anchors. Moreover, practical localization requires reliable CSI reception from all anchors across the entire target area, which is difficult to guarantee in the Marshall Student Center due to its large and highly dynamic environment.

Instead, the localization experiments were conducted in a conference hall of size $22\,\text{m} \times 17\,\text{m}$, shown in Fig.~\ref{fig:trilateration_map}, with three ESP32 access points acting as anchors and one ESP32 station acting as the receiver. CSI data were collected at 98 different points distributed approximately uniformly over the localization region. These points were arranged on a rectangular grid with 14 points along one dimension and 7 points along the other, yielding a total of $14 \times 7 = 98$ measurement locations. This grid-based collection ensured that the receiver was sampled over the full region of interest while maintaining a controlled 2D geometry. All devices were kept at approximately the same height so that the problem remained consistent with 2D localization.

For localization, three separate fully connected networks were trained, one per anchor, using raw CSI to estimate receiver-to-anchor distance. The three predicted distances were then combined through deterministic trilateration to estimate 2D receiver coordinates. To form the train--test split, 80\% of the grid points were used for training and the remaining 20\% were selected uniformly at random for testing. For each anchor, the regressor was trained for 50 epochs with batch size 64 and learning rate $10^{-3}$ on the same workstation used in the authentication experiments.

Table~\ref{table:rmse-trilateration} shows that the distance regressors achieve sub-meter to near-meter accuracy for the individual anchors: 0.696\,m for A1, 1.029\,m for A2, and 0.847\,m for A3. The variation reflects differences in propagation conditions such as geometry, multipath, and local obstructions. When these estimates are combined through trilateration, the final localization RMSE is 1.180\,m. Although slightly higher than the best individual anchor errors, this is expected because trilateration accumulates errors from all three anchors. Overall, the results show that CSI contains useful distance-related information and that the proposed framework can achieve practical meter-level indoor localization under realistic dynamic conditions.

\section{Conclusion}

In this work, we studied CSI-based radio frequency fingerprinting in a realistic and highly dynamic indoor environment using low-cost ESP32 devices. We proposed \textit{Contra}, a contrastive learning framework that maps CSI samples and device identities into a shared embedding space for transmitter authentication. Across multiple encoder backbones, the results showed that grouping consecutive CSI packets improves authentication performance, and that preserving the spatiotemporal structure of CSI is important for robust learning under crowded, multipath-rich conditions. Among the evaluated models, the ViT-based ContraCSI variant achieved the best closed-set authentication performance.

We also extended the framework to open-set authentication by combining the embedding space learned by Lite3D-CNN-ContraCSI with a GEM--CUSUM sequential detector. The resulting verifier--classifier pipeline achieved strong rejection of unseen transmitters while maintaining low false positive rates, showing that contrastive CSI embeddings are useful not only for closed-set identification but also for open-set verification. As an additional extension, we investigated CSI-based indoor localization through distance regression and trilateration, and obtained practical meter-level localization accuracy in a realistic indoor environment.

Overall, the results demonstrate that CSI-based RFF can remain effective in dynamic real-world environments when both the data collection setup and the learning framework are designed to account for channel variability. In future work, we will conduct similar dynamic-environment experiments using chaotic antenna arrays (CAA), building on our previous hardware-based RF fingerprinting studies, in order to assess whether the enhanced signatures of CAA can provide further gains in robustness and authentication performance under realistic crowded indoor conditions.

\printbibliography
\end{document}